\documentclass[12pt]{article}
\usepackage{graphicx}
\usepackage[T2A]{fontenc}
\usepackage[utf8]{inputenc}
\usepackage[russian,english]{babel}
\usepackage{amsfonts}
\usepackage{amssymb}
\usepackage{amsmath}
\usepackage{pgfplots}
\usepackage{braket}
\usepackage{tikz}
\usepackage{float}
\newcommand{\w}{\omega}
\newcommand{\s}{\sigma}
\newcommand{\la}{\lambda}
\renewcommand{\a}{\alpha}
\begin{document} 
\title{Quality of Control in the Tavis-Cummings-Hubbard Model}

\author{Raffael D\"ull$^1$, Alexei Kulagin$^2$, Wanshun Lee$^2$,  \\
Yuri Ozhigov$^{2,3}$, Miao Huei-huei$^2$, Zheng Keli$^2$,\\
{\it 
1. Technical University Munich, Munich, Germany.}\\
{\it 2. Moscow State University of M.V.Lomonosov, Faculty VMK}
\\
{\it Moscow 119992, Vorobyovy Gory, GSP-2, 2nd hum. corpus, SKI.}
\\
{\it 3. K.A.Valiev Institute of physics and technology RAS, Moscow, Russia}\\
{\it Nakhimovsky Prospekt 32-a}
\\ 
}
\maketitle

PACS: 03.65,  87.10\\
Keywords: Optical field, Tavis-Cummings-Hubbard model, Quantum gate, Decoherence

\begin{abstract}
The quality of controlling a system of optical cavities in the Tavis-Cummings-Hubbard (TCH) model is estimated with the examples of quantum gates, quantum walks on graphs, and of the detection of singlet states. This type of control of complex systems is important for quantum computing, for the optical interpretation of mechanical movements, and for quantum cryptography, where singlet states of photons and charges play an essential role. It has been found that the main reason for the decrease of the control quality in the THC model is due to the finite width of the atomic spectral lines, which is itself related to the time energy uncertainty relation. This paper evaluates the quality of a CSign-type quantum gate based on asynchronous atomic excitations and on the optical interpretation of the motion of a free particle.
\end{abstract}

\section{Introduction}

The Tavis-Cummings-Hubbard model TCH (\cite{JC}, \cite{Tav}) in quantum electrodynamics is well suited to simulate the dynamics of small atomic ensembles that interact with light. The high flexibility of the model allows us to explore both single-photon quantum scenarios (\cite{Su}, \cite{Su1}) and multi-photon scenarios (\cite{Sup1}, \cite{Sup}). In this work, we will analyze and evaluate dynamic quantum scenarios with one or two photons, which are suitable to simulate the dynamics of a broad range of quantum systems. In many cases, it is much easier to manipulate the dynamics of photons than of charges. We will call this technique {\it optical interpretation} and demonstrate it on three examples: the motion of a free massive particle, the entanglement of a quantum gate and the optical selection of dark states in atomic ensembles. These tasks are important to develop optimal technology for quantum computers and for quantum cryptography applications. The limitations of all three scenarios within the framework of the TCH model will be discussed both with theoretical considerations and with computer simulations. 

Even standard quantum key distribution protocols that operate on individual photons and do not use entanglement, such as B84 or BB92, provide a fundamentally new level of secrecy compared to classical protocols. However, using singlet states of both photons ans massive particles provide additional advantages as in protocol AK-2017 (\cite{Mat}). The key challenge in this protocol is to prepare singlet photon (or atoms) states in the Bell state of the form $\frac{1}{\sqrt 2}(|01\rangle-10\rangle)$.

The preparation of singlet atomic states and of optical imitations is limited by the technical capabilities of the physical implementation of the TCH model. The quality of quantum gates for computations depends on the theoretical boundaries of the TCH model itself.

The Tavis-Cummings Hamiltonian describes a system of $s$ identical two-level atoms in an optical resonator. The transition energy between the two levels matches with high accuracy the energy $\hbar\w$ of a photon trapped in the optical cavity. The interaction energies between the atoms and the field within the cavity is denoted by $g_j,\ j=1,2,...,s$. If $g_j/\hbar\w\ll 1$, the rotating wave approximation can be used and the Hamiltonian takes the form: 
\begin{equation}
	H_{TC} = \hbar\w(a^+a + \sum\limits_{j=1}^s \s^+_j\s_j) + a^+\bar\s + a\bar\s^+
\end{equation}
where $\bar\s=\sum\limits_{j=1}^sg_j\s_j$. The field operators for photon creation $a^+$ and annihilation $a$ depend on the given cavity, so an additional subscript $q$ is provided to identify the respective cavity. Similarly, the atomic operators $\s_j^+$ and $s_j$ denote the excitation and relaxation of the $j$-th atom in the cavity.

Cavities can be connected with atoms tuned to the same frequency $\w$ to form a network of $c$ cavities in which photons can move between two cavities $q$ and $p$ with an amplitude $\nu_{q,p} = r_{q,p} e^{i\phi_{q,p}}$. The parameters $r_{q,p}$ and $\phi_{q,p}$ depend linearly on the thickness of and length of the optical fiber connecting the involved cavities. Such a system is described by the Hamiltonian:
\begin{equation}
	H_{TCH}=\sum\limits_{q=1}^cH^q_{TC}+\sum\limits_{1\leq q<p\leq c}\nu_{q,p}a_q^+a_p + \bar\nu_{q,p}a_qa_p^+
\end{equation}
Note that it is always possible to choose the waveguide length such that the phase of the transition amplitudes between the cavities is zero, thus the amplitude $\nu_{q,p}$ remains real and only its absolute value may change.

\section{Optical interpretation of the dynamics of a free massive particle}

For now, we consider a system of connected optical cavities without any atoms, such that one photon is located in the quantum state $|\Psi(x,t)\rangle$, where $x$ is the index of the cavity and $t$ is the current time. The task is to select waveguides in a way that the dynamics of this state can be approximated by a free particle in a one-dimensional space, whose coordinate is proportional to the cavity index $x$.

This problem is the simplest example of a continuous-time quantum walk (CTQW) \cite{Am}. This kind of dynamics allows for far-reaching generalizations, such as the introduction of a magnetic field as suggested by \cite{2L}, where the motion of a free particle was analyzed in a two-dimensional space. We will compare the representation of the dynamics of a one-dimensional free particle in the JCH model without atoms with a simple analytical expression of the Feynman kernel. 

All physical quantities are represented as qubits, for which they are defined as arithmetic progressions on a finite set of values. After an appropriating linear transformation ${\cal D}$ on the coordinate $x$, which is equivalent to choosing a new unit of length, we can assume that the definition domain of the wave function $[x_0,x_N]$ coincides with the interval $[0,1]$. For $N=2^n$ quantities, we then obtain $x_q=q/N$ for $\ q=0,1,...,N-1$ and we approximate the coordinate $x$ with accuracy $1/N$ by the by the binary string $|e_1e_2...e_n\rangle$ where the binary digits $e_k\in\{ 0,1\}$ come from the expansion $q=\sum\limits_{k=1}^{n}2^{n-k}e_k$.

Any of such strings represents a basis state of a system with $n$ qubits, which can then be used as the discrete representation of the wave function. This representation does not have a physical dimension, since only the transition operator ${\cal D}$ from the continuous function $\Psi(x)$ to the qubit representation and the basis states $|j\rangle$ are dimensional. The coordinate of the variable $a$ is $a/\sqrt{N}$ and lies within the interval $[0,\sqrt{N}]$ and its unit could be the Planck constant in the proper system of units. The term $c/\sqrt{N}$ has then to be associated with a momentum, and it seems natural to assume that it belongs to the interval $[-\sqrt{N}/2,\sqrt{N}/2]$ because a particle in $[0,\sqrt{N}]$ can move in both directions. Therefore, the momentum has to be equal to $\sqrt{N}(c/N-1/2)$.

The discrete form of the momentum operator is the $N$-dimensional Hermitian operator
\begin{equation*}
	p_{discr}=QFT\sqrt{N}(x_{discr}-I/2)QFT^{-1}=A^{-1}QFT\sqrt{N}x_{discr}QFT^{-1}\ A
\end{equation*}
 with the diagonal operator $A=diag(e^{\pi i a})_{a=0,1,...,N-1}$. Its eigenvectors are given by $A^{-1}QFT^{-1}|a\rangle$ in \eqref{Fourier} and are associated to the respective eigenvalues $\sqrt{N}(a-1/2)$ for all $\ a=0,1/N,...,(N-1)/N$.

The dynamics of a free particle with nonzero mass is determined by the Hamiltonian $H_{free}=p^2/2m$, which is expressed in the coordinate basis as $H_{free}=QFTdiag(p^2/2m)QFT^{-1}$. The Quantum Fourier Transforms are given by: 
\begin{equation}
\begin{array}{ll}
\quad QFT: |c\rangle&\rightarrow \frac{1}{\sqrt N}\sum\limits_{a=0}^{N-1}e^{-\frac{2\pi i ac}{N}}|a\rangle\\
QFT^{-1}: |a\rangle&\rightarrow \frac{1}{\sqrt N}\sum\limits_{c=0}^{N-1}e^{\frac{2\pi i ac}{N}}|c\rangle\end{array}
\label{Fourier}
\end{equation}

Each value of the coordinates $x_q=q/N$ for the indices $ q=0,1,...,N-1$ is associated to one optical cavity without atoms, which are connected by optical fibers such that the transition amplitude $\nu_{q,p}$ is equal to the term $\langle q|H_{free}|p\rangle$ dependent on the Hamiltonian $H_{free}$ of a free particle. Such a network of $N$ connected cavities
stands for the motion of a free particle along a straight line.

\begin{figure}[H]
\centering
\includegraphics[width=0.95\textwidth]{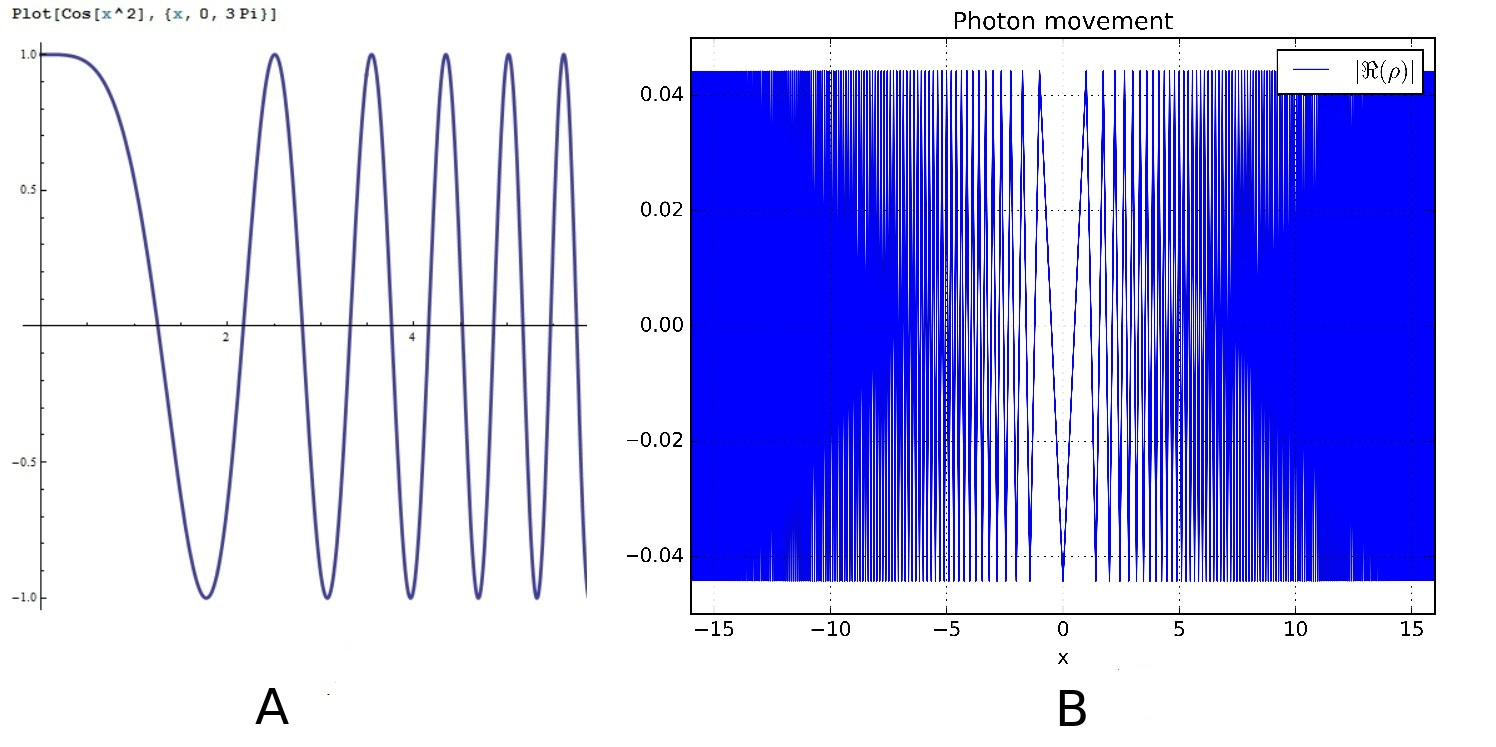} 
\caption{{\bf A. } General aspect of the real part of the Feynman kernel $K(x,t)$ for a free particle with nonzero mass, which is located at the initial time $t=0$ at the initial position $x=0$. {\bf B.} Real part of the amplitude of the photon position in the optical network simulating a free particle. The phase shift is due to the choice of the initial photon position ($x=0$) in the cavity system at the point with coordinate $\sqrt{N}/2$. The scale along the axes has arbitrary units.}
\label{fig:4}
\end{figure}

The quality criterion is determined by the Feynman kernel of a free particle of the form:
\begin{equation}
\label{fey}
K(x,t)=At^{-\frac{1}{2}}e^{\frac{imx^2}{\hbar t}};
\end{equation}
Its real part is shown in figure \ref{fig:4}, А.

\begin{figure}
\centering
\includegraphics[scale=0.08]{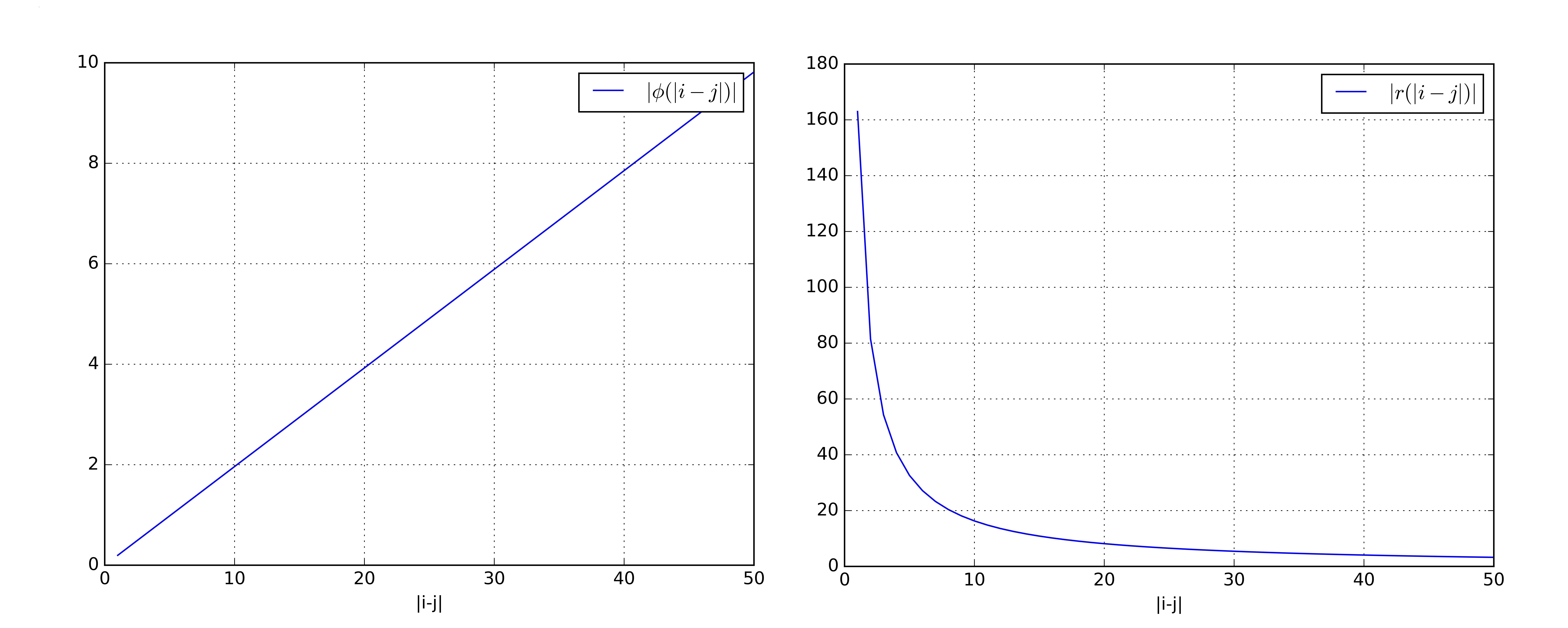} 
\caption{Dependence of the phase and the transition amplitude on the distance between points along the propagation line for the Hamiltonian JCH, which simulates the propagation of a free particle}
\label{fig:4.1}
\end{figure}

Figure \ref{fig:4.1} shows how the phase and the transition amplitude between cavities depend on the distance between the respective points on the straight line along which the particles propagate. The length of the waveguide is proportional to the real distance from the initial particle position and its thickness is inversely proportional to this distance. The calculation was preformed on a personal computer and the number of optical cavities did not exceed a couple of hundreds.

As it can be seen in Figure \ref{fig:4}, the dynamics of a massive particle can only be reproduced with low quality, such an optical interpretation can thus only simulate the general aspect of the dynamics. The advantage of the model is that it uses the same basic means as quantum gates, therefore the free movement of massive particles in space can be easily included in a quantum computer based on the TCH model, without the need to solve the Cauchy problem of the Schr\"odinger equation, as in the standard approach for quantum computers \cite{Za}.

\section{Quality of an entaglement gate on optical cavities}

The entanglement gate plays a key role for quantum computing (see \cite{Fe}, \cite{G},\cite{Sh}) because, together with one-qubit gates, it allows to implement every imaginable quantum circuit as described in \cite{universal}.

Photonic systems are one of the most common ways to construct such gates, as for instance with the well-known KLM-scheme \cite{KLM} or by improving \cite{DC} with quantum teleportation \cite{tele} and by switching on atomic excitation \cite{Po}. The results of experiments on optical cavities in \cite{Re} or \cite{Wi} indicate the practical feasibility of such gates. 

The CNOT gate is a common example of entanglement gates, it acts on a pair of qubits $|x,y\rangle$ according to the rule $CNOT:\ |x,y\rangle\rightarrow |x,x\oplus y\rangle$. In  \cite{Wi}, it was implemented using the vibrational degrees of freedom of an atom. It is equivalent to the well-known universal gate CSign, which acts according to the rule $CSign:\ |x,y\rangle\rightarrow (-1)^{xy}|x,y\rangle$, in the sense that one of these gates can be obtained from the other with the help of simply realizable one-qubit gates: $CNOT=w_y\ CSign\ w_y$, where the Hadamard gate $w_y = (\s_x+\s_y)/\sqrt 2$ is applied to the second, controlled qubit of the operator argument. The CSign gate inverts the phase of the basis state $|11\rangle$. The closely related operator $coCSign:\ |x,y\rangle\rightarrow (-1)^{x(y\oplus 1)}|x,y\rangle$ can easily be reduced to the CSign operator with the help of a one-qubit gate: $coCSign=\s_x(y) CSign\ \s_x(y)$. 

The implementation of the coCSign gate is proposed in \cite{Oz}. It partially uses the idea of \cite{A}, where the states of dual-rail photons are used as qubits, however the scheme in \cite{Oz} is much simpler. The present paper discusses the quality of this implementation of the gate.

The focus of this paper is therefore on the coCSign gate $coCSign:\ |x,y\rangle\rightarrow (-1)^{(x\oplus 1)y} |x,y\rangle$. Since it is related to $CSign=\sigma_x(x)\ coCSign\ \sigma_x(x)$ by single qubit-gates, which are implemented as linear optical devices, the presented estimate also applies to the $CNOT$ gate.

\section{Layout of the coCSign gate}
We now consider a cavity with one atom inside whose energy gap between the ground state $|0\rangle$ and the excited state $|1\rangle$ exactly matches the energy $\hbar\w$ of the confined photon in the cavity. Under the assumption of the rotating wave approximation, the Hamiltonian takes the form:
\begin{equation}
H=H_{JC}=H_0+H_{int};\ H_0=\hbar\omega a^+a+\hbar\omega\sigma^+\sigma,\ H_{int}=g(\a^+\sigma+a\sigma^+),
\label{HamJC}
\end{equation}
The basic states of the atom and of the field can be written as $|n\rangle_{ph}|m\rangle_{at}$, where $n=0,1,2,...$ is the number of photons in the cavity and $m=0,1$ is the level of atomic excitation. In our setup, there are in total 2 photons, so $n$ can take the values $n=0,1,2$. To implement  the gate, a term $H_{jump}$ is added to the Hamiltonian above to represent the transition of a photon from the cavity $i$ to $j$ and vice versa. 
\begin{equation}
\label{nu}
H_{jump}=\nu (a_ia^+_j+a_ja^+_i), 
\end{equation}

Let $\tau_1=\pi\hbar/g,\ \tau_2=\pi\hbar/g\sqrt{2}$ be the periods of the Rabi oscillations for the total energies $\hbar\omega$ and $2\hbar\omega$. The operator $U_t=e^{-\frac{i}{\hbar}Ht}$, that describes the evolution of the system for most of the time, depends on the total energy in the cavity. If this energy is equal to $\hbar\omega$ in the basis states $|\phi_0\rangle=|1\rangle_{ph}|0\rangle_{at}$ and $ |\phi_1\rangle=|0\rangle_{ph}|1\rangle_{at}$, we obtain:
\begin{equation}
U_{\tau_1/2}=-i\sigma_x,\ U_{\tau_1}=-I,\ U_{2\tau_1}=I,
\label{phase_add}
\end{equation}
In this equation, $\sigma_x$ is the Pauli matrix. A similar relation holds for $\tau_2$ for the total energy of the cavity $2\hbar\omega$. The phase of the state is relevant here, since states with different numbers of photons in the cavities may be in superposition. 

The transition of a photon from a cavity $i$ to another cavity $j$ is carried out by the simultaneous switching on of the Pockels cells, which translates into the addition of the term $H_{jump}$ to the main interaction $H_{int}$. In the absence of atoms, it leads to the same dynamics as in Rabi oscillations, but with a period $\tau_{jump}=\pi\hbar/\nu$. In the ideal case, we can assume that $\nu\gg g$, which means that it is possible to move a photon between two cavities without being affected by the atom at all. The paper \cite{A} describes the difficulty to implement this behavior in an experiment, however, we will consider this problem as a technical issue.

Because of the incommensurability of the Rabi oscillation periods $\tau_1$ and $\tau_2$, we can choose two natural numbers $n_1$ and $n_2$ such that the following approximate equality holds.
\begin{equation}
2n_2\tau_2\approx 2n_1\tau_1+\frac{\tau_1}{2}
\label{noncommon}
\end{equation}
The nonlinear shift required to implement the $coCSing$ gate is obtained with exactly this difference in the period.

The basis state $|0\rangle$ is created in our model as the state of the optical cavity $|0\rangle_{ph}|1\rangle_{at}$, and the basis state $|1\rangle$ as $|1\rangle_{ph}|0\rangle_{at}$. Thus, the two-qubit state $|01\rangle$ on which the phase shift should act, represents the state $|01\rangle_{ph}|10\rangle_{at}$, where the first photonic qubit belongs to the cavity $x$ and the second to the cavity $y$. It is to be noted that after a time $\tau_1/2$, the null and zero states change place with a phase advance of $-\pi/2$. 

\begin{figure}[H]
\centering
\includegraphics[scale=0.60]{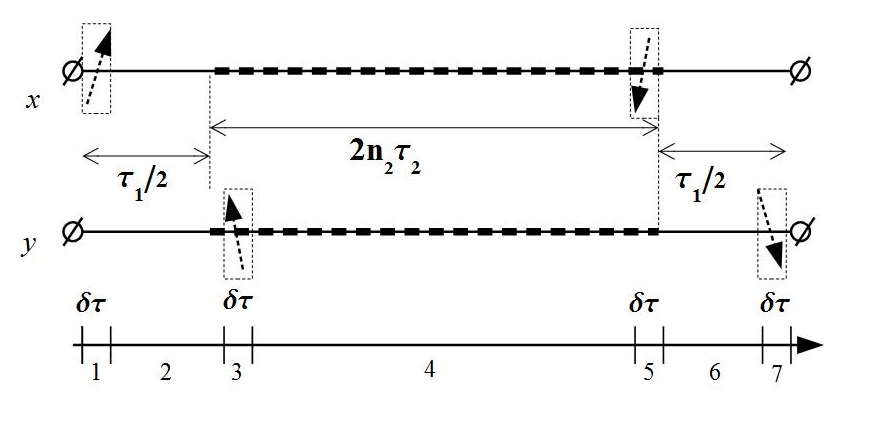}
\caption{Operation sequence of the implementation of the coCSign gate: $|x,y\rangle\rightarrow (-1)^{(x\oplus 1)y}|x,y\rangle$ on asynchronous atomic excitations in optical cavities, split into 7 time segments, and short photon transition times $\delta\tau=\tau_{jump}/2\ll\tau_{1(2)}$. The expected behavior of the coCSign gate is observed after an additional $\tau_1/2$ at the end of diagram}
\label{fig:1}
\end{figure}

\begin{figure}[H]
\centering
\includegraphics[scale=0.60]{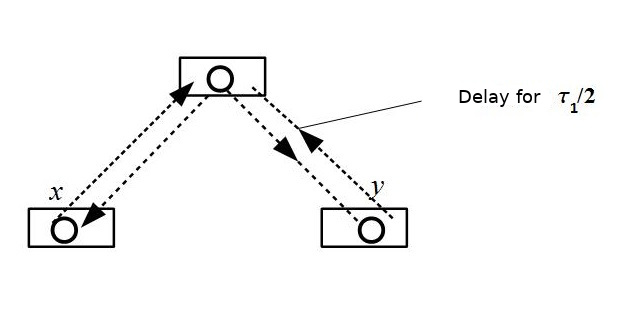}
\caption{The three cavities ($x,y$ and auxiliary) involved in the coCSign gate}
\label{fig:coCSign}
\end{figure}

The sequence of operations to implement the $coCSign$ gate is shown in Figure \ref{fig:1} and the involved cavities are shown in Figure \ref{fig:coCSign}. In the beginning, an almost instantaneous photon exchange between the auxiliary and the $x$ cavity is induced. After a delay $\tau_1 / 2$, the same photon exchange happens between the auxiliary and the $y$ cavities. Then, the system evolves without any exchanges for the time $2n_2\tau_2$, before the same sequence of photon exchanges occurs again between the auxiliary and respectively the $x$ and $y$ cavities with a time gap $\tau_1/2$ in between. With this choice of photon displacement times, it follows that at these moments there are exactly one or zero photons in a cavity. Therefore, switching on the Pockels cells during a short time $\delta \tau=\pi\hbar/2\nu\ll \tau_1$ results in exactly the desired photon movements.

In the work \cite{Oz}, it was proved that this scheme ideally gives an implementation of the coCSign gate. In the article \cite{SOL}, it was calculated that even if the numbers $n_1$ and $n_2$ are below 100, and if all other decoherence factors are ignored, the gate accuracy exceeds 90\%. 

The interaction force between the atom and the field in the cavity takes the form:
\begin{equation}
g=\sqrt{\hbar\omega/V}d\ E(x),
\label{g}
\end{equation}
where $V$ is the effective volume of the cavity, $d$ is the dipole moment of the transition between the ground an the perturbed states and $E(x)$ describes the spatial arrangement of the atom in the cavity and is equal to $E(x)=sin(\pi x/L)$ with $L$ the length of the cavity. To ensure the confinement of the photon in the cavity, $L$ has to be chosen such that $L=n\la/2$ is a multiple of the photon wavelength $\la$. In experiments, $n=1$ is often chosen to decrease the effective volume of the cavity, which makes it possible to obtain dozens of Rabi oscillations (see, for example \cite{Re}).

The physical limitation on the gate quality follows from the energy-time uncertainty relation for photons. The parameter $\nu$ for the photon transition from one cavity to another cannot be chosen too large because a short transition time automatically means a large uncertainty in its energy. A large error in the energy results in an altered wavelength and if the wavelength becomes too different from twice the cavity length, the photon can escape the cavity. 

For instance, the photon frequency in experiments with the rubidium atom is approximately $10^{10}\ sec^{-1}$ and the upper bound for the possible uncertainty in frequency is $10^9\ sec^{-1}$ (the real bound is however much lower). Taking into account the uncertainty relation $\delta\omega\ \delta t\approx 1$, a lower estimate for the photon transition time window could be given by $\delta\tau\approx 10^{-9}\ sec$. Since the period of the Rabi oscillations is approximately $\tau\approx 10^{-6}\ sec$, we obtain the bounds $10^{-9}\leq\delta\tau\ll 10^{-6}$ for the transition time, which means that a single execution of the gate will induce an error of a least than $10^{-3}$.



\begin{figure}[H]
	\centering
	\includegraphics[scale=0.055]{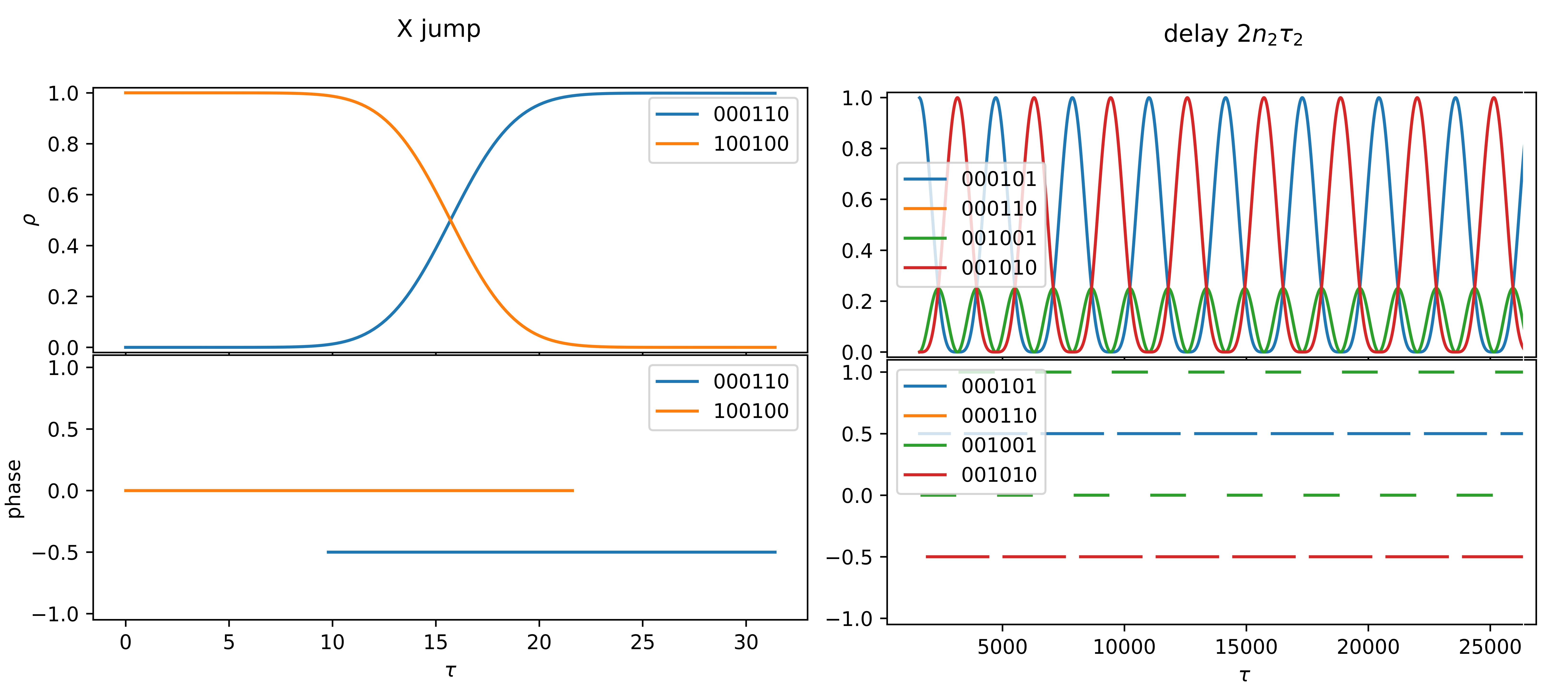} 
	\caption{Left side: exchange of a photon between the auxiliary and $x$ cavities for the ground-state atoms, with the parameter $\alpha=\tau_1/2 \sqrt{2\pi}\sigma$. Right side: Evolution of the segment with length $2n_2\tau_2$. }
	\label{fig:j}
\end{figure}

Let us suppose, that the intensity $\nu$ of any photon transfer between cavities from equation \eqref{nu} is described by the Gaussian
\begin{equation}
\label{nu_}
\nu(t)=\alpha e^{-\frac{(t-t_{id})^2}{2\sigma^2}}.
\end{equation}

The parameter $\sigma$ needs to be chosen $\sigma\ll\tau_{1,2}$ to fulfill the aforementioned condition for the gate implementation. In this case, the transition of a photon from the cavity $x$ to the auxiliary cavity and back is regular for the ground states of an atom and an example of it is shown in the left part of Figure \ref{fig:j}. The evolution of the state in the central segment of duration $2n_2\tau_2$ is shown in the right part.

\begin{figure}[H]
\centering
\includegraphics[scale=0.5]{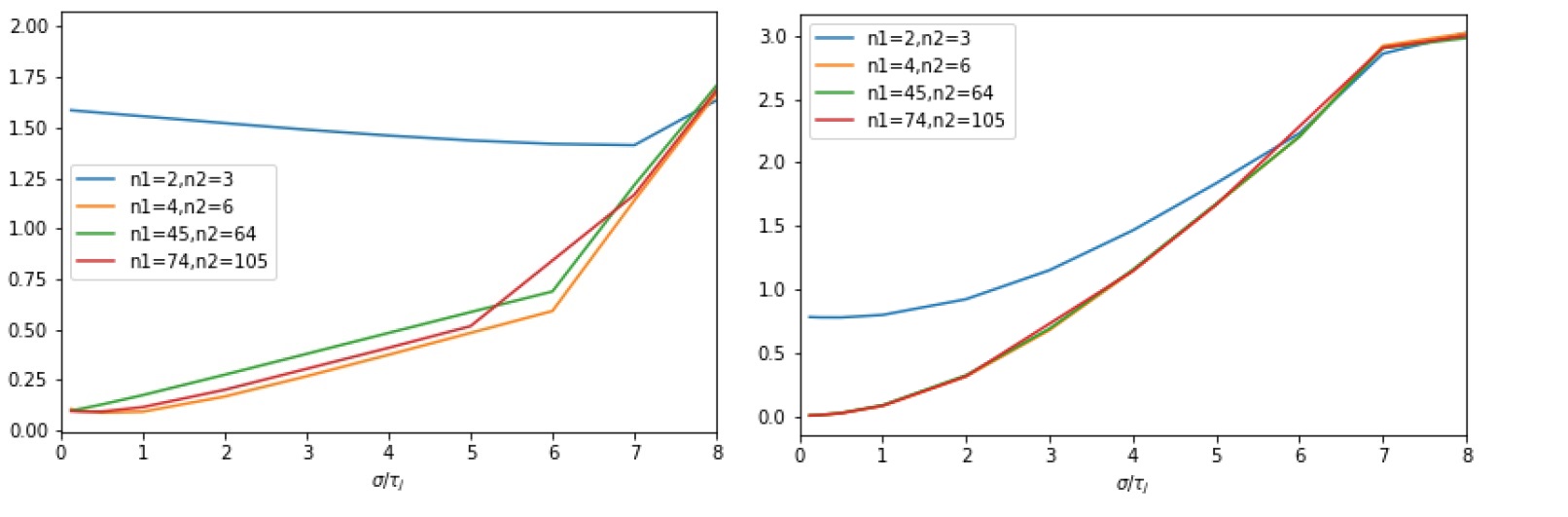} 
\caption{Trace and modular distances between the ideal and the decoherent results of the coCSign gate or the initial state $|\psi_0\rangle$ and for various combinations of $n_1$ and $n_2$.}
\label{fig:11}
\end{figure}

To assess the quality of the gate, we consider the initial state $|\psi_0\rangle = \frac{1}{2}(|00\rangle+|01\rangle+|10\rangle+|11\rangle)$. First, we assume that the maximum intensity $\alpha$ has a fixed value. In this case, the accuracy can be evaluated by the trace and modular distances between the ideal state $\rho_{id}=|\Psi_{id}\rangle\langle\Psi_{id}|$ and the real state $\rho=|\Psi\rangle\langle\Psi|$ of the system. We define the differences $\delta\rho=\rho-\rho_{id}$ and $\delta\Psi=|\Psi\rangle-|\Psi_{id}\rangle$ to estimate the errors:
\begin{equation}
d_{tr}=tr\sqrt{\delta\rho^+\delta\rho} \qquad \ d_{mod}=\delta\Psi^+\delta\Psi.
\end{equation}
The results of the numerical simulations is shown in Figure \ref{fig:11}.

\begin{figure}[H]
\centering
\includegraphics[scale=0.43]{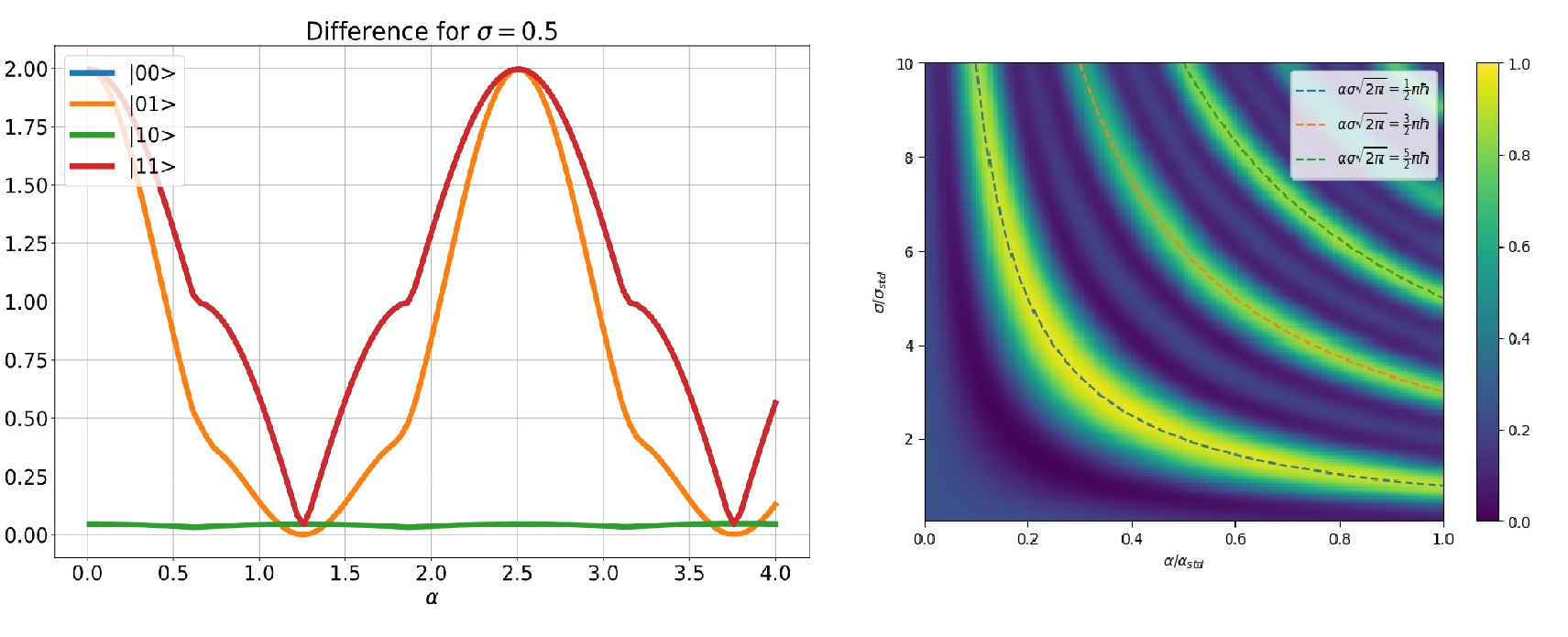} 
\caption{{\bf A.} Modular distance between the ideal and decoherent states after the application of the coCSign gate on different initial states in function of $\alpha$ for the fixed values $\sigma=0.5, g=10^{-3},\hbar=1,\w=1$. {\bf B.} Modular error for variable  $\alpha$ and $\sigma$. }
\label{fig:2}
\end{figure}

Figure \ref{fig:2} shows the modular distance between the ideal and the simulated, decoherent gate for a fixed $\sigma$ and variable $\alpha$ and $\sigma$. The observed periodicity follows from the possibility of multiple photon transitions between the cavities during one activation of the Pockels cells for increasing $\alpha$.

\section{Dark states of two-level atomic ensembles}

The complexity of the model can be further increased if we consider several atoms in one cavity. A pure quantum effect then occurs: dark states of atomic ensembles.

In a dark state, an atom cannot emit any light, although its total energy is not zero. For two-level atoms, such states are obtained as the linear combinations of tensor products of singlets. With the RWA approximation, the ground states of atom parts are also allowed in the product. The Bell state $|s\rangle=\frac{1}{\sqrt 2}(|01\rangle-|10\rangle)$ has the remarkable property that for any unitary operator $U$, its tensor square does not change the singlet state: $U^{\otimes 2}|s\rangle=|s\rangle$. The quantum protocol to distribute the secret key AK-2017 is based on this singlet property \cite{Mat}.

The preparation of such singlet states of photons is therefore a key issue for quantum cryptographic protocols, and the quality assessment of this preparation plays a practical role. Singlets can be generalized to ensembles of multi-level atoms by taking the determinant of their base states. For instance, the multi-singlet for a three-level atom takes the form $|D_3\rangle=\frac{1}{\sqrt 6}\sum\limits_{\pi\in S_3}(-1)^{\s(\pi)}|\pi(1)\pi(2)\pi(3)\rangle$, where the sum is taken over all permutations of three atoms. 

The simplest way to obtain dark singlet atomic states is through optical selection. One photon is launched into a cavity with an atomic ensemble and a vacuum state of the field. Then, its escape time from the cavity is measured, and after many repetitions, the average photon emission time can be  determined very accurately by virtue of the central limit theorem. If the initial state of the atomic ensemble is a superposition of singlet products (for a diatomic ensemble, of a single singlet), the photon escapes much faster because it cannot interact with the atoms. On the other hand, if the ensemble is not in a singlet state, there is interaction between the photon and the atoms, which causes a delay in its exit. 

\begin{figure}[H]
	\centering
	\includegraphics[scale=0.6]{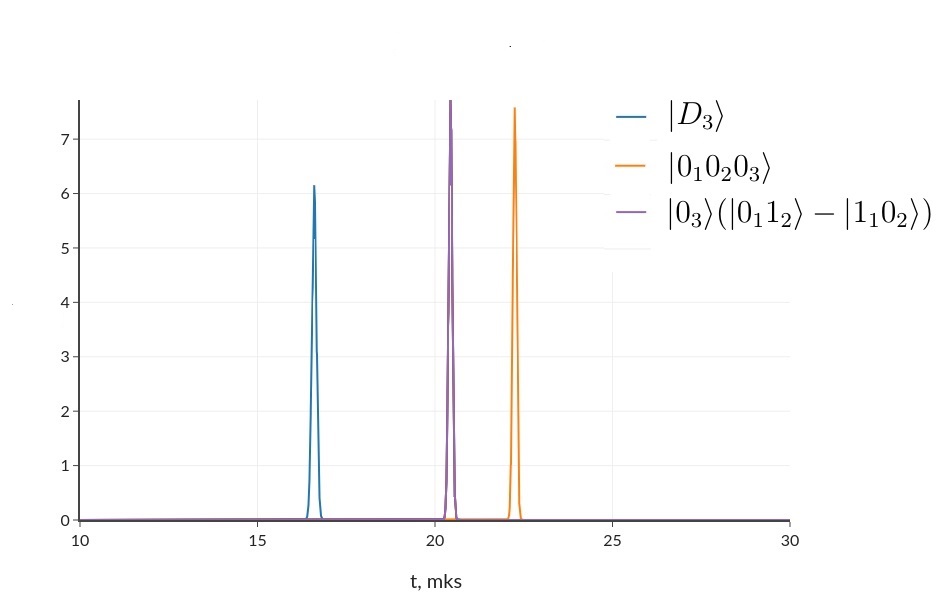} 
	\caption{Distribution density of the photon emission time from the cavity for one dark and two different types of light states of an atomic ensemble with three $Rb^{85}$ atoms (Figure taken from \cite{KO}).}
	\label{fig:10}
\end{figure}

The optical selection makes it also possible to estimate the degree of darkness of an atomic ensemble, as illustrated by Figure \ref{fig:10}.

In \cite{KO}, it was shown that the quality of optical selection is reduced due to the quality of the photon detectors, where the error of any kind is less then 3\%. The accuracy of the selection itself is much higher, since this systematic error does not affect the result obtained by the comparison of two alternative hypotheses of dark or light states in the ensemble. The random error decreases rapidly with the increase in the number of experiments. 

\section{Conclusion}
The quality assessment of quantum control with the Tavis-cummings-Hubbard model was carried out on the standard problem of a free quantum particle, on the main controlled two-qubit gate in quantum computing and on the problem of obtaining dark states of two-level atoms, which is important for quantum cryptography. Decoherence occurs because of the incommensurability of the Rabi oscillation periods and because of the broadening of spectral lines for an atom in a cavity due to the energy-time uncertainty relation. These effects, in combination with the finite response time of the Pockels cell, may significantly reduce the accuracy of the gate. The latter factor seems to be only a technical issue, and was not further investigated in this work. The incommensurability induces an error of the order of one percent, which is a very good result for the gate. However, this is not a definite assessment of the feasible gate quality. The broadening of spectral lines was taken into account too roughly to draw certain conclusions about its contribution to the error and needs a more detailed analysis. Together with the technical limitations on the response rate of the Pockels cell, it imposes significant restrictions on the achievable accuracy of the main entanglement gate, and this regardless of its specific shape (CNOT, CSign or coCSign), and yields a serious obstacle to scale the Feynman scheme to a quantum computed based on gates.

It was also shown that for dark states of atomic ensembles, quantum operations can be achieved with a significantly higher accuracy, even with relatively imprecise photon detectors. This is facilitated by statistical methods to choose between two alternative hypotheses, however these are only measurable operations. The use of statistical methods in quantum computing means elements of classical parallelization, similar to the KLM scheme \cite{KLM}, which should be part of any implementation of a quantum computer. 

Dark states can theoretically persist for a very long time, since the atomic ensemble in such a state does not interact with the resonator field. Therefore, the development of the concept of quantum computing from the traditional Feynman scheme to the use of dark states promises new weapons in the fight against decoherence. 

The analyzed examples demonstrate the great importance of the Tavis-Cummings-Hubbard model in theoretical estimates of the quality of quantum operations. This model could be the basis for the creation of an operating system for a quantum computer, where the central part of the computing technology bases on the first principles of quantum computing.

\section{Acknowledgments}
This work was supported by the Russian Foundation for Basic Research, grant a-18-01-00695.

\end{document}